\documentstyle[psfig,11pt]{article}
\begin{document}
\vspace{1.0cm}
{\Large \bf THE BROAD BAND X-RAY SPECTRA OF INTERMEDIATE\\
            BL LAC OBJECTS}

\vspace{1.0cm}

J. Siebert$^1$,  W. Brinkmann$^1$, S.A. Laurent-Muehleisen$^2$, and M. Matsuoka$^3$

\vspace{1.0cm}
$^1${\it Max-Planck-Institut f\"ur extraterrestrische Physik, Postfach 1603,
    85740 Garching, Germany}\\
$^2${\it IGPP/LLNL, 7000 East Ave., Livermore, CA 94550, USA}\\
$^3${\it RIKEN, Hirosawa 2-1, Wako, Saitama 351-01, Japan}\\

\vspace{0.5cm}

\section*{ABSTRACT}
We present recent X-ray observations of intermediate BL Lac objects (IBLs), 
i.e. BL Lacs which are located between high-energy and low-energy peaked 
BL Lac objects with respect to $\alpha_{\rm rx}$. We briefly 
discuss the statistical properties of IBLs from the RGB sample and then
focus on a detailed broad band spectral analysis of two objects, namely 
1424+2401 and 1055+5644, which were observed with ASCA and SAX, respectively.
In both cases the spectra are steep and we find significant curvature in
1424+2401 in the sense that the spectrum gets flatter at both low ($< 1$ keV) and 
high energies ($>$ 5 keV). Our results are in line with the hypothesis of
a continuous distribution of synchrotron peak frequencies among BL Lac objects.

\section{INTRODUCTION}
The spectral energy distribution (SED) of BL Lac objects is dominated
by non-thermal emission from a relativistic jet oriented close
to the line of sight. It is characterised by two peaks, where
the first lies in the IR to soft X-ray range and is thought to be
due to synchrotron radiation. The second peak is in the $\gamma$-ray
range and most likely due to inverse--Compton radiation.

It is well established that BL Lac objects detected in radio
surveys (RBLs) show markedly different properties compared to
BL Lacs detected in X-ray surveys (XBLs). In general, the latter
are less extreme in terms of variability, polarization, superluminal
motion and luminosity. Also, RBLs show stronger optical emission lines
than XBLs and most of the BL Lac objects detected in $\gamma$--rays
belong to the RBL class.

The unification of RBLs and XBLs is still a matter of debate. The
angle between the relativistic  jet and the line of sight is
certainly a crucial parameter (e.g. Celotti et al. 1993), but it has
been shown (e.g. Sambruna et al. 1996) that orientation alone cannot
explain the large differences in the spectral energy distribution of
the two classes. A second parameter could for
example be the jet kinetic luminosity (Georgantopoulos \& Marscher 1998) or
the frequency of the synchrotron peak (e.g. Padovani \& Giommi 1995). The
latter suggestion has led to a re-classification of BL Lacs
into HBLs and LBLs, i.e. high-- and low--frequency peaked BL Lac objects,
where most of the XBLs and RBLs belong to the HBL and LBL class, respectively.

Selection effects, caused by the high flux limits of previous
samples (e.g. 1-Jy sample, EMSS sample), clearly are important and
might explain much of the apparent dichotomy of BL Lac objects.
Therefore it is desirable to establish new, large and complete samples
of BL Lac objects with deeper flux limits. In particular, the properties
of intermediate BL Lacs (IBLs), i.e. those which fill the gap between
LBLs and HBLs, play an important role for BL Lac unification theories.

\section{THE RGB SAMPLE OF INTERMEDIATE BL LAC OBJECTS}

Recently, a new large sample of BL Lac objects has become available
(Laurent-Muehleisen et al. 1998). It is a subsample of the so-called 
RGB sample, which resulted from the cross-correlation of the ROSAT 
All-Sky Survey and the 87GB 5GHz radio survey (Brinkmann et al. 1995, 1997). 
After accurate radio positions were obtained by VLA observations of more than
1900 sources of this sample (Laurent-Muehleisen et al. 1997), the optical
identification of a properly selected subgroup resulted in more than 50
new BL Lacs, adding to the about 100 previously known BL Lac objects
contained in the RGB sample.

The distribution in $\alpha_{\rm rx}$, the spectral index between 5GHz and 
1 keV, is shown in Fig.~1 for the RGB BL Lacs. For comparison the distributions 
for the 1-Jy and the EMSS sample are also shown. Clearly, there is a continuous 
distribution in $\alpha_{\rm rx}$ and the gap between HBLs and LBLs is filled 
with a number of objects. There are 22 BL Lacs with 0.7$< \alpha_{\rm rx} <$0.8, 
which constitute our sample of intermediate BL Lac objects.

\begin{figure}
\hspace*{1.2cm}
\begin{minipage}{7cm}
\psfig{figure=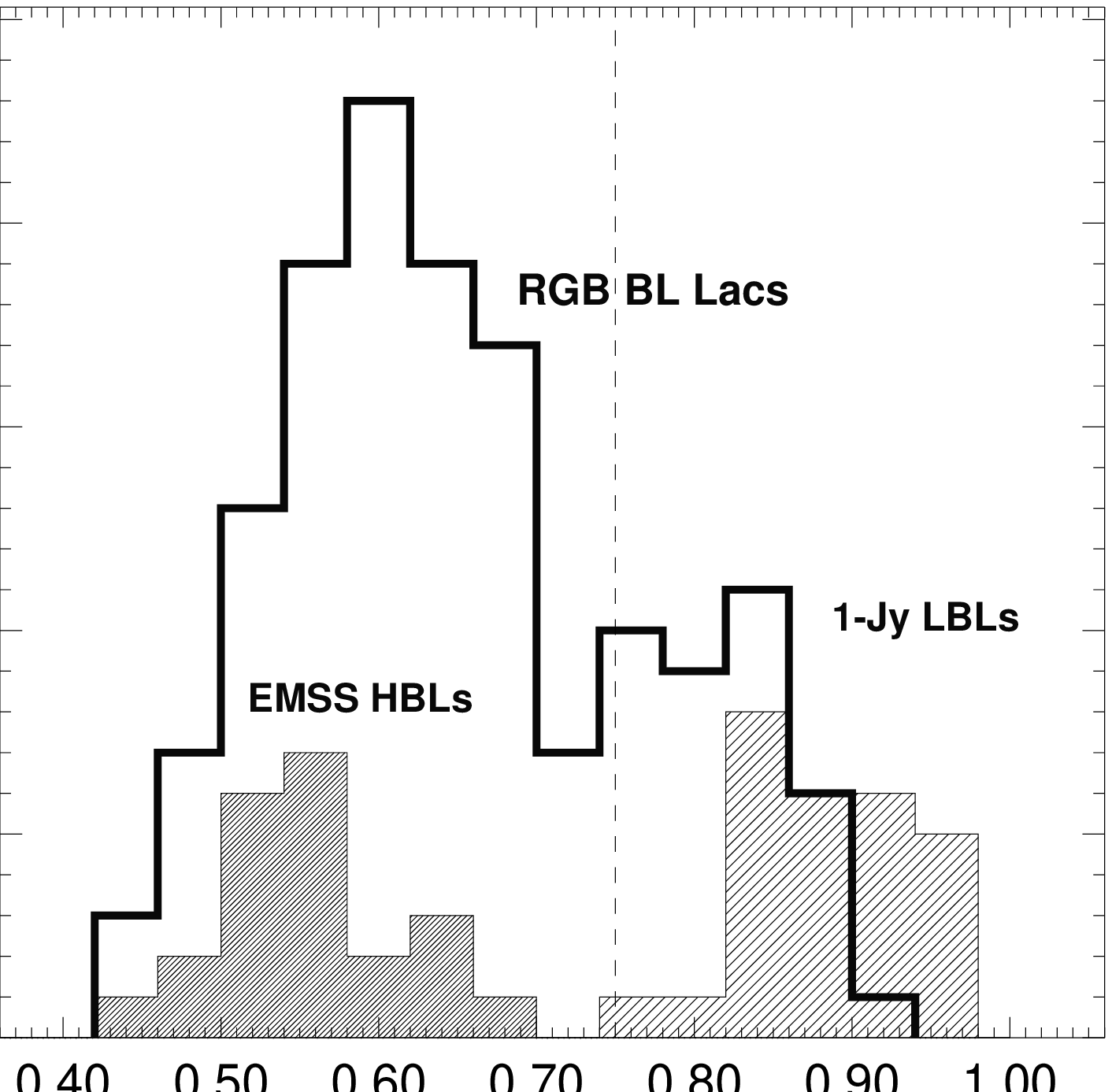,width=7cm}
\caption{The distribution of $\alpha_{\rm rx}$ for HBLs, LBLs and the BL Lacs
from the RGB sample. $\alpha_{\rm rx}$ is the two-point spectral index between 5 GHz
and 1 keV. The dashed line indicates the canonical dividing line between HBLs and
LBLs.}
\end{minipage}
\hspace*{0.8cm}
\begin{minipage}{7cm}
\vspace*{0.8cm}
\psfig{figure=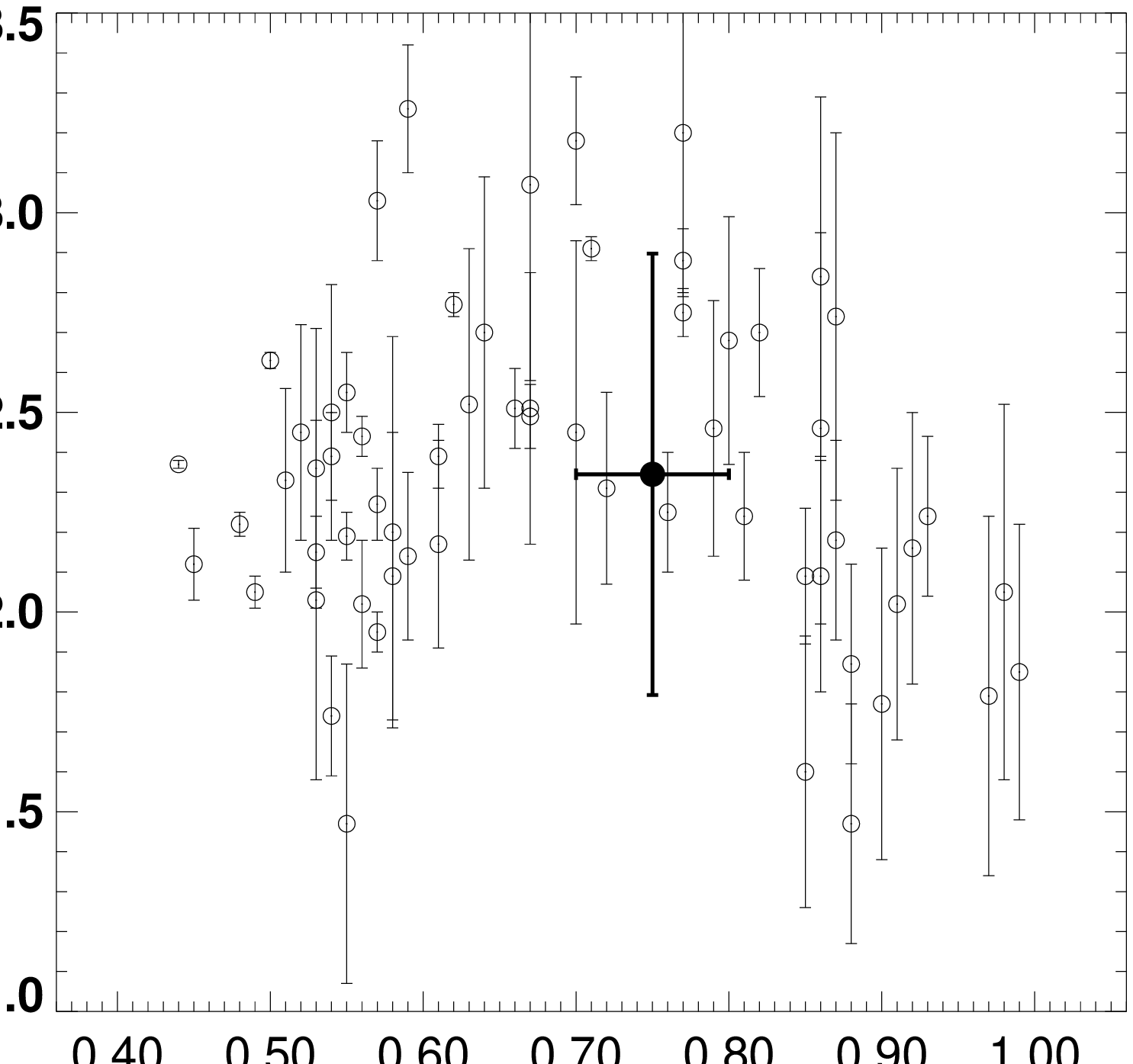,width=7cm}
\caption{ROSAT PSPC photon index $\Gamma$ versus the two point spectral index
$\alpha_{\rm rx}$ between 5 GHz and 1 keV. The open circles denote the BL Lac objects
discussed in Lamer et al. (1996). The filled circle represents the average
$\Gamma$ for the 22 IBLs of the RGB sample and the error bar corresponds to the
standard deviation.}
\end{minipage}
\end{figure}

Fig.~2 shows the soft X-ray photon index $\Gamma$ in the 0.1--2.4 keV energy
band as a function of $\alpha_{\rm rx}$. The data are taken from Lamer et al. 
(1996), who derived the X-ray spectral shape of 74 BL Lac objects from pointed
PSPC observations. The photon indices for the IBL sample were determined from the
hardness ratios. The average value as well as the standard deviation are
plotted in Fig.~2 (filled circle). The positive and negative correlations for HBls
and LBLs, respectively, are predicted if the population of BL Lac objects is
indeed characterised by a continuous distribution of peak frequencies (Lamer et al. 
1996). IBLs should exhibit the steepest X-ray spectra in this scenario and our
average value ($\langle\Gamma_{\rm IBL}\rangle = 2.35\pm0.55$) is roughly 
consistent with this prediction.  

\section{THE BROAD BAND X-RAY SPECTRUM OF TWO IBLS}

\subsection{\underline{1424+2401}}

\begin{figure}
\hspace*{1.2cm}
\begin{minipage}{7cm}
\psfig{figure=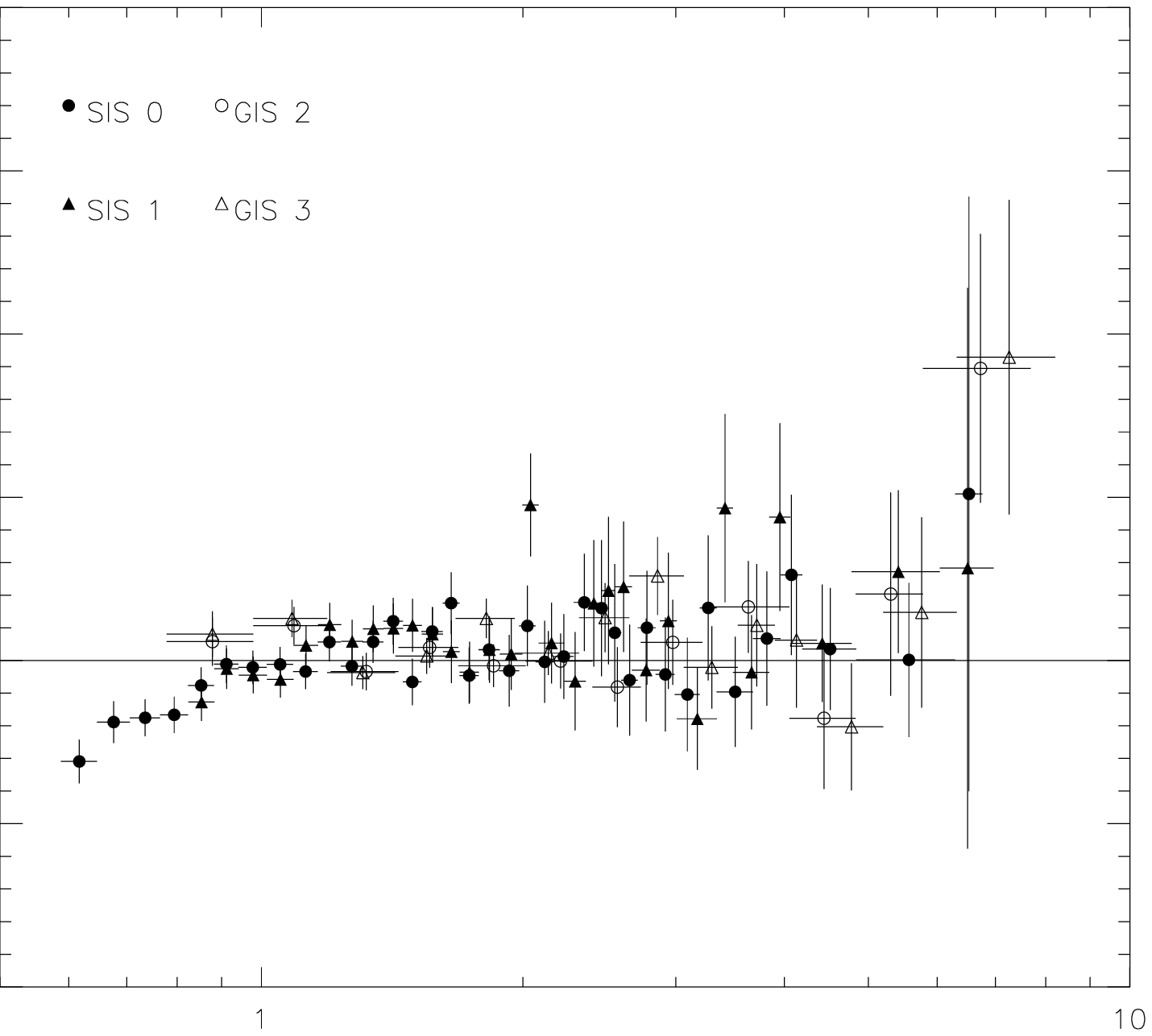,width=7cm}
\caption{Ratio of a simple power law model to the ASCA data from all four
detectors. The photon index, determined between 2 and 5 keV is
$\Gamma = 2.82\pm0.16$. Note the flattening of the spectrum at low and high
energies.}
\end{minipage}
\hspace*{0.8cm}
\begin{minipage}{7cm}
\psfig{figure=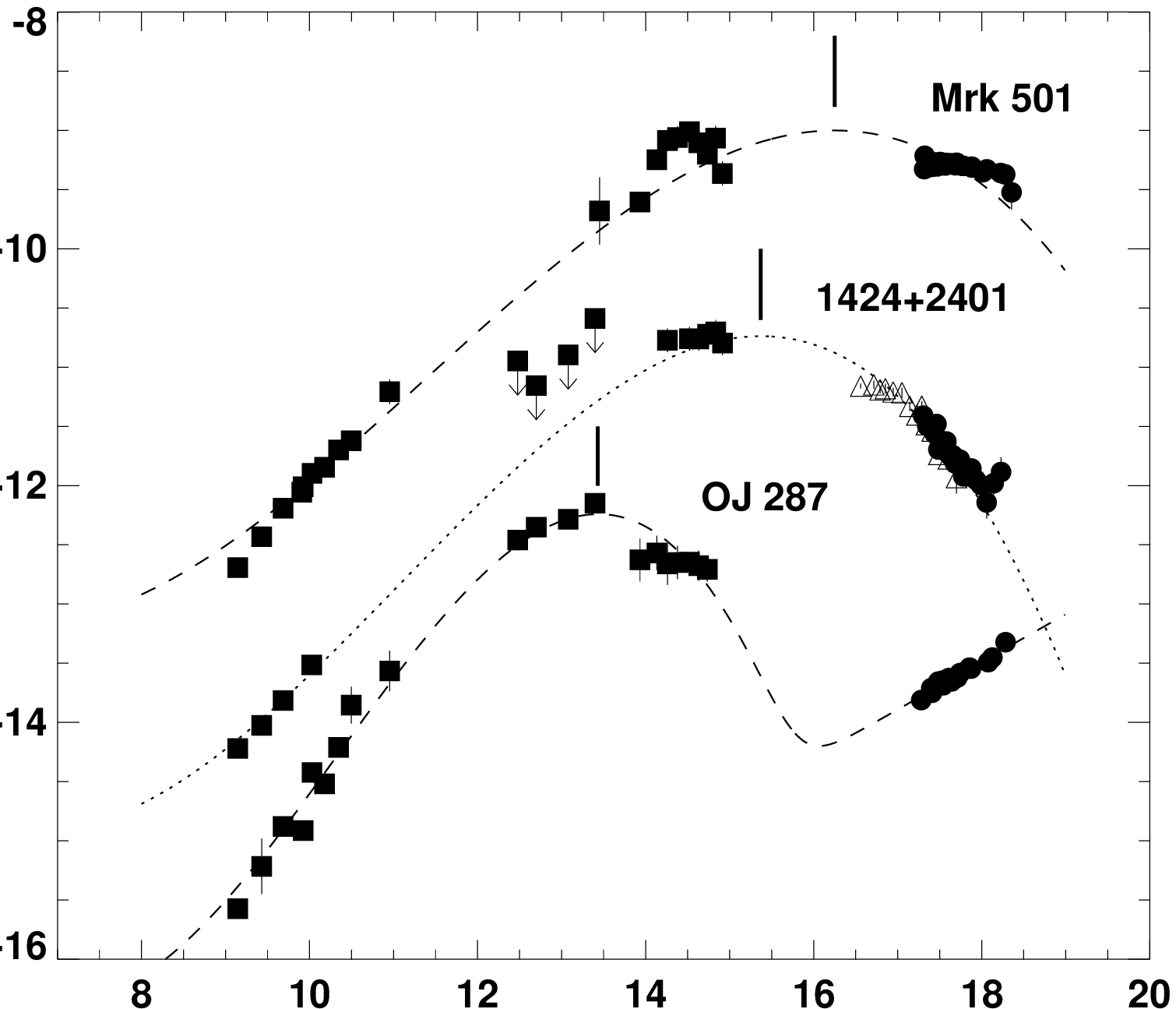,width=7cm}
\caption{The spectral energy distribution of 1424+2401 in comparison to
Mrk 501 (HBL) and OJ 287 (LBL). The vertical scaling is arbitrary and the vertical
bars denote the frequency of the synchrotron peak.}
\end{minipage}
\end{figure}

1424+2401 was originally considered to be a white dwarf, but was re-classified
as a BL Lac by Impey \& Tapia (1988) and Fleming (1993). A representation of the 
ASCA spectrum of this source is shown in Fig.~3. The ratio of the spectral data 
to a simple power law model with a photon index of $\Gamma = 2.82\pm0.16$, which
was determined in the energy range from 2 to 5 keV, shows systematic residuals. 
The spectrum gets flatter at the low and the high energy end and a model with two 
energy breaks around 1 and 5 keV significantly improves the fit. We interpret
this result as evidence for a energy dependent curvature in the ASCA spectrum of
1424+2401.  
 
The harder spectrum towards higher energies might be interpreted in terms of a flat
inverse--Compton component that starts to dominate the X-ray spectrum. The high
energy photon index is only weakly constrained by the ASCA data
($\Gamma_{\rm E>5 keV} = 0.97_{-0.77}^{+0.98}$), but is completely consistent with
the observed high energy spectra of LBLs ($\Gamma\sim 1.6$), which are thought to
be dominated by IC emission. The flux deficit at lower energies cannot be due to 
absorption, since the implied N$_{\rm H}$ values would be inconsistent with the 
ROSAT PSPC spectrum of 1424+2401.

The spectral energy distribution SED of 1424+2401 is compared to Mrk\,501 and
OJ\,287 in Fig.~4. The latter represent typical HBLs and LBLs, respectively. The
vertical bars denote the peak frequencies of a third-order polynomial
fit to the datapoints (dashed lines) and the open triangles represent the
absorption corrected ROSAT PSPC spectrum of 1424+2401. The peak frequency and the 
very steep X-ray spectrum confirm the intermediate nature of 1424+2401 and hence 
indicate a continuous distribution of peak frequencies in BL Lac objects.

\subsection{\underline{1055+5644}}

1055+5644 is one of the newly identified BL Lac objects from the RGB sample.
The combined SAX LECS/MECS data of 1055+5644 again indicate a steep X-ray
spectrum ($\Gamma = 2.46\pm0.20$). However, no significant curvature is seen, apart
from a marginal flattening in the spectrum above 5 keV in the MECS data (Fig.~5).
In this context we note that 1055+5644 is a $\gamma$-ray source listed in the 
third EGRET catalog (Hartmann et al., this volume), which implies that the
flattening seen in the SAX data might be real. The SED of 1055+5644, shown in Fig.~6,
is again typical for IBLs, i.e. it shows a peak frequency in the optical/NIR.

\begin{figure}
\hspace*{1.2cm}
\begin{minipage}{7cm}
\hspace*{-0.5cm}
\psfig{figure=siebert-fig5.ps,width=7.5cm,height=6.cm,angle=-90}
\caption{Data, model and residuals (bottom panel) of a simple power law model fit to the 
SAX LECS and MECS data of 1055+5644. The best-fitting photon index is 
$\Gamma = 2.46\pm0.20$.}
\end{minipage}
\hspace*{0.8cm}
\begin{minipage}{7cm}
\psfig{figure=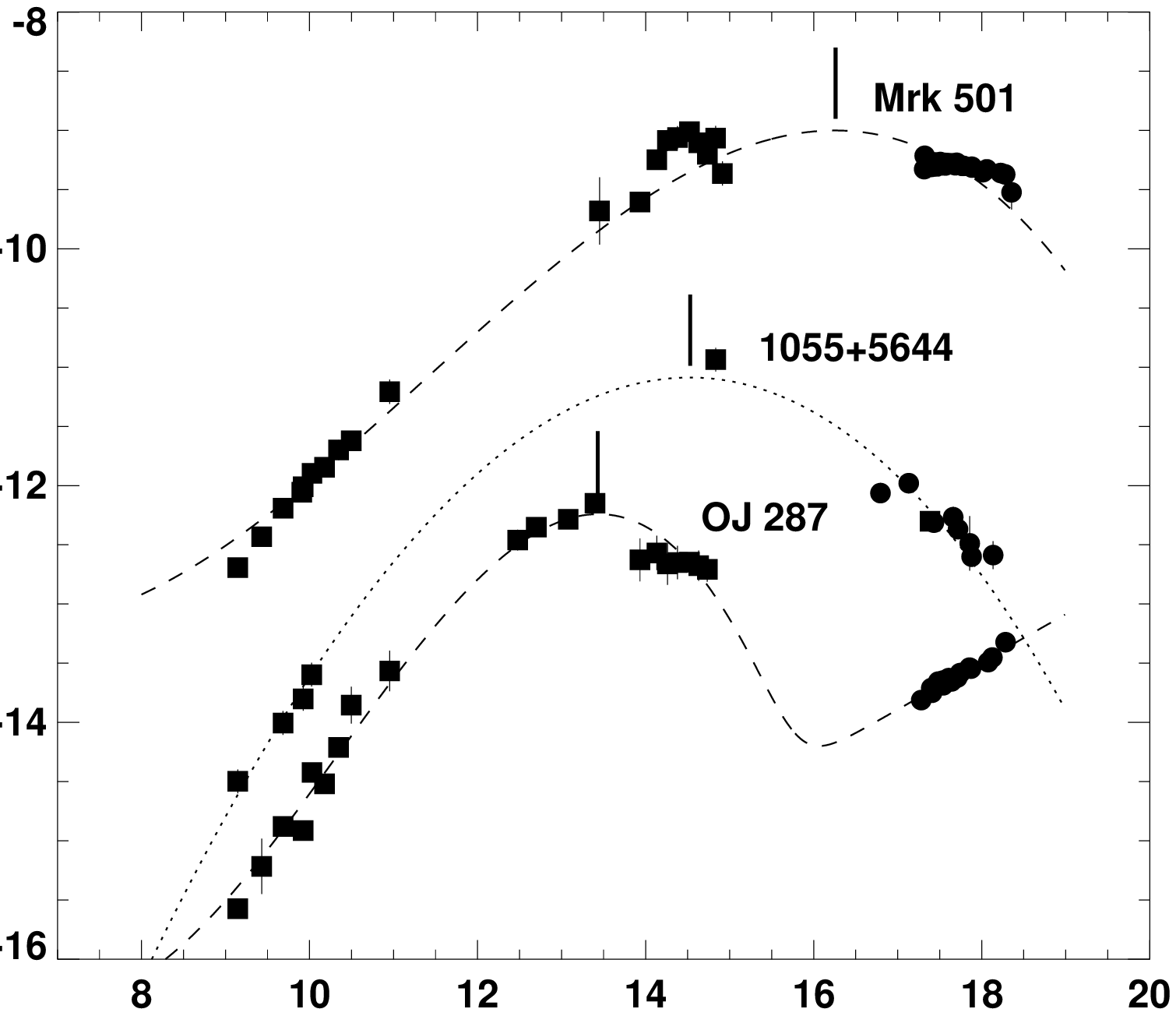,width=7cm}
\caption{The spectral energy distribution of 1055+5644 in comparison to
Mrk 501 (HBL) and OJ 287 (LBL). The vertical scaling is arbitrary and the vertical
bars denote the frequency of the synchrotron peak.}
\end{minipage}
\end{figure}

\section{REFERENCES}
\vspace{-5mm}
\begin{itemize}
\setlength{\itemindent}{-8mm}
\setlength{\itemsep}{-1mm}

\item[]
Brinkmann, W., Siebert, J., Reich, W., F\"urst, E., Reich, P., Voges, W., Tr\"umper, J.,
Wielebinski, R., {\bf A\&AS}, 109, 147 (1995)  

\item[]
Brinkmann, W., Siebert, J., Feigelson, E.D., Kollgaard, R.I., Laurent-Muehleisen, S.A.,
McMahon, R.G., Reich, W., F\"urst, E., Reich, P., Voges, W., Tr\"umper, J., {\bf A\&A}, 323,
739, (1997)

\item[]
Celotti, A., Maraschi, L., Ghisellini, G., Caccianiga, A., Maccacaro, T.,
ApJ, {\bf 416}, 118 (1993)

\item[] 
Fleming, T.A., Green, R.F., Januzzi, B.T., Liebert, J., Smith, P.S., Fink,
H.H., {\bf AJ}, 106, 1729 (1993)

\item[]
Georganopoulos, M., Marscher, A.P., {\bf ApJ}, in press

\item[]
Impey, C.D., Tapia, S., {\bf ApJ}, 333, 666 (1988)

\item[]
Lamer G., Brunner H., Staubert R., {\bf A\&A}, 311, 384 (1996)

\item[]
Laurent-Muehleisen, S.A., Kollgaard, R.I., Ryan, P.J., Feigelson, E.D.,
Brinkmann, W., Siebert, J., {\bf A\&AS}, 122, 235 (1997)

\item[]
Laurent-Muehleisen, S.A., Kollgaard, R.I., Ciardullo, R., Feigelson, E.D.,
Brinkmann, W., Siebert, J., {\bf ApJS}, in press (1998)

\item[]
Padovani, P., Giommi, P., {\bf ApJ}, 444, 567 (1995)




\end{itemize}

\end{document}